\documentclass[sigconf,natbib=true,camera-ready]{acmart}
\copyrightyear{2025}
\acmYear{2025}

\acmConference[SIGIR '25]{the 48th International ACM SIGIR Conference on Research and Development in Information Retrieval}{July 13-18, 2025}{Padua, Italy}

\setcctype{by}

\AtBeginDocument{%
  }

\copyrightyear{2025}
\acmYear{2025}
\setcopyright{cc}
\setcctype{by}
\acmConference[MM '25]{Proceedings of the 33rd ACM International Conference
on Multimedia}{October 27--31, 2025}{Dublin, Ireland}
\acmBooktitle{Proceedings of the 33rd ACM International Conference on
Multimedia (MM '25), October 27--31, 2025, Dublin,
Ireland}\acmDOI{10.1145/3746027.3758310}
\acmISBN{979-8-4007-2035-2/2025/10}

\usepackage[most]{tcolorbox}
\usepackage{tabularx}
\usepackage{bbm}
\usepackage{hyperref} 
\usepackage{multirow}
\usepackage{makecell}

\usepackage{amssymb}
\usepackage{pifont}
\newcommand{\cmark}{\ding{51}}%
\newcommand{\xmark}{\ding{55}}%
\begin{document}

\title{Small Stickers, Big Meanings: A Multilingual Sticker Semantic Understanding Dataset with a Gamified Approach}

\author{Heng Er Metilda Chee}
\email{xxe23@mails.tsinghua.edu.cn}
\affiliation{%
\institution{DCST, Tsinghua University}
   \city{}
  \country{}
\institution{Beijing, China}
   \city{}
  \country{}
\institution{Quan Cheng Laboratory, Jinan, China}
   \city{}
  \country{}
}

\author{Jiayin Wang}
\email{JiayinWangTHU@gmail.com}
\affiliation{%
  \institution{DCST, Tsinghua University}
   \city{Beijing}
 \country{China}
}



\author{Zhiqiang Guo}
\email{georgeguo.gzq.cn@gmail.com}
\affiliation{%
  \institution{DCST, Tsinghua University}
   \city{Beijing}
 \country{China}
}

\author{Weizhi Ma}
\authornote{Corresponding authors. 
\\This work is supported by the Natural Science Foundation of China (Grant No. U21B2026, 62372260). Weizhi Ma is also sponsored by Beijing Nova Program.
}
\email{mawz@tsinghua.edu.cn}
\affiliation{%
\institution{AIR, Tsinghua University}
 \city{Beijing}
 \country{China}}



 
\author{Min Zhang}
\authornotemark[1]
\email{z-m@tsinghua.edu.cn}
\affiliation{%
\institution{DCST, Tsinghua University}
   \city{}
  \country{}
\institution{Beijing, China}
   \city{}
  \country{}
\institution{Quan Cheng Laboratory, Jinan, China}
   \city{}
  \country{}
  }


\makeatletter
\gdef\@copyrightpermission{
  \begin{minipage}{0.2\columnwidth}
   \href{https://creativecommons.org/licenses/by/4.0/}{\includegraphics[width=0.90\textwidth]{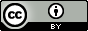}}
  \end{minipage}\hfill
  \begin{minipage}{0.8\columnwidth}
   \href{https://creativecommons.org/licenses/by/4.0/}{This work is licensed under a Creative Commons Attribution International 4.0 License.}
  \end{minipage}
  \vspace{5pt}
}
\makeatother

\renewcommand{\shortauthors}{Heng Er Metilda Chee, Jiayin Wang, Zhiqiang Guo, Weizhi Ma \& Min Zhang}


\begin{abstract}

Stickers, though small, are a highly condensed form of visual expression, ubiquitous across messaging platforms and embraced by diverse cultures, genders, and age groups. Despite their popularity, sticker retrieval remains an underexplored task due to the significant human effort and subjectivity involved in constructing high-quality  query-sticker datasets. Although large language models (LLMs) excel at general NLP tasks, they falter when confronted with the nuanced, intangible, and highly specific nature of sticker query generation.

To address the challenge of collecting diverse and contextually appropriate sticker search queries, we introduce Sticktionary, a gamified annotation framework designed to elicit high-quality, semantically rich queries from contributors. Using this framework, we construct StickerQueries, a multilingual dataset comprising 1,115 English and 615 Chinese sticker search queries, annotated by over 60 contributors across more than 60 hours of annotation. We demonstrate the utility of StickerQueries in several downstream tasks, including query generation and sticker retrieval. Through comprehensive quantitative and qualitative evaluations, we demonstrate that StickerQueries significantly improves the quality of query generation, retrieval accuracy, and semantic understanding within the sticker domain. To support future research, we publicly release the multilingual StickerQueries dataset and two fine-tuned query generation models\footnote{https://huggingface.co/datasets/metchee/sticker-queries}. Additional experiments and supplementary case studies can be found here \footnote{https://arxiv.org/abs/2506.01668}.

\end{abstract}




\maketitle

\section{Introduction}
\label{sec:intro}

\begin{figure}[h]
    \centering
    \includegraphics[width=0.87\linewidth]
    {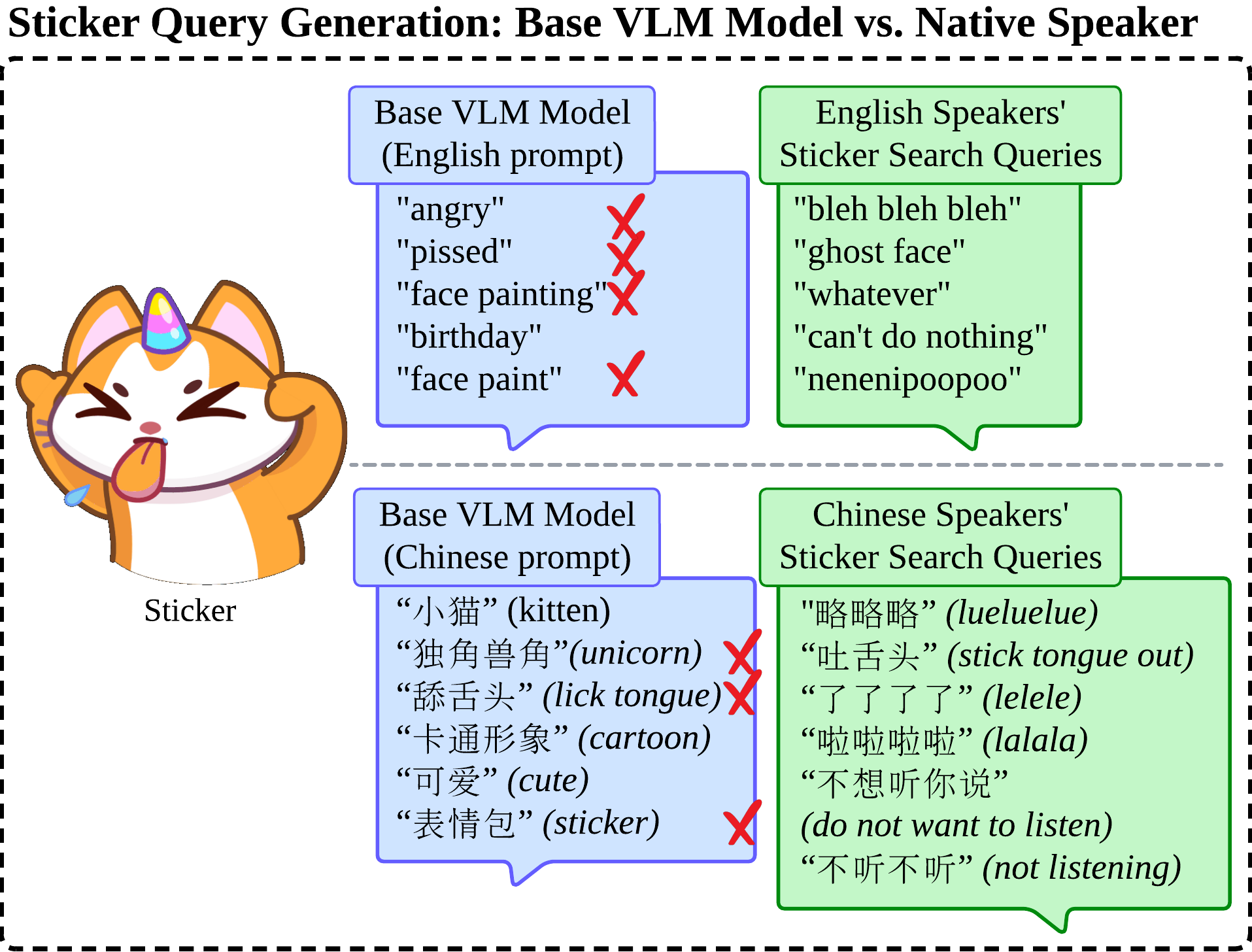}
    \caption{Base VLM models are unable to understand and generate accurate sticker search queries, which requires depth and precision, oftentimes hallucinating. This persists in both English and Chinese languages.}
    \label{fig:sticker_semantic_understanding}
\end{figure}
Stickers have become an integral component of modern digital communication. Widely used across diverse demographics—spanning genders, cultures, and age groups—they enrich messaging experiences by conveying emotions, reactions, and humor in a compact, expressive format. While several sticker datasets have been developed, few resources focus on the task of sticker retrieval, despite its centrality in real-world applications. This gap is largely due to the intensive human effort required to create high-quality datasets, and the challenges of interpreting and formalizing the rich semantics embedded in sticker use.

Despite their small size, stickers are capable of conveying deeply nuanced human emotions. For instance, expressing mild annoyance, frustration, or sarcasm might be easily achieved with a single sticker (e.g., Fig.~\ref{fig:sticker_semantic_understanding}), but would require a verbose textual explanation. Sticker queries often involve a rich interplay of modalities—such as sound (e.g., “bleh bleh bleh”), action (“ghost face”), emotion (“whatever”), culture (“nenenipoopoo”), thought (“can’t do nothing”), and tone. These phenomena are observed across both Chinese and English contexts.
Generating such queries requires cultural sensitivity, contextual understanding, and linguistic precision—capabilities that even advanced vision-language models (VLMs) struggle to replicate. Sticker query generation is prone to hallucinations, and typical models often fail to grasp the subtle social and emotional layers stickers encode.

To address these limitations, we introduce StickerQueries, a multilingual sticker query dataset for understanding and retrieving stickers in context. We summarize our contributions in the following:

\begin{enumerate}
\item We propose Sticktionary, a gamified annotation framework that encourages contributors to create expressive, context-rich sticker queries through an engaging game-based interface. This design fosters the generation of high-quality and semantically resonant queries.
\item We construct and release StickerQueries, a dataset consisting of 1,115 English and 615 Chinese queries, annotated by over 60 contributors with more than 60 hours of effort. We also release two fine-tuned query generation models on Hugging Face.
\item Through extensive experiments, we demonstrate significant improvements in sticker retrieval and query generation, validating our dataset and methodology.
\end{enumerate}

\section{Related Work}
\label{sec:related_work}

\begin{table*}[htbp]
\centering
\small
\caption{Comparison of sticker datasets sorted by characteristics and year. Currently, no sticker query dataset has been publicly available except our proposed StickerQueries. We use NA as a shorthand for Not Available data values.}
\begin{tabular}{lcllccc}
\toprule
\textbf{Category} & \textbf{Dataset} & \textbf{Year} & \textbf{Description} & \textbf{Publicly Available?} & \textbf{Query Language?} & \textbf{Native EN?} \\ \toprule

\textit{Dialogues} & SRS, PESRS \cite{learning-to-respond-2021, learning-to-respond-with-stickers-2020} & 2020 & Sticker dialogues. & \xmark  & \xmark & \xmark \\
 & MOD \cite{MOD} & 2021 & Sticker dialogues & \cmark  & \xmark & \xmark \\
 & StickerInt \cite{stickerint} & 2024 & Sticker dialogues & \xmark  & \xmark & \xmark \\
 & MCDSCS \cite{mcdscs} & 2024 & Sticker dialogues. & \cmark & \xmark & \xmark \\
 & STICKERCONV\cite{zhang2024stickerconv} & 2024 & Simulated sticker dialogues. & \cmark & \xmark & \xmark \\
 & U-Sticker \cite{chee2025106kmultitopicmultilingualconversational} & 2025 & Continuous sticker dialogues. & \cmark & \xmark & \xmark \\
 & MultiChat \cite{multichat} & 2025 & Sticker dialogues. & \cmark  & \xmark & \xmark \\ \midrule

\textit{Sentiment} & CSMSA \cite{CSMSA} & 2022 & Sticker sentiment. & \xmark  & \xmark & \xmark \\
 & Sticker820K \cite{stickerclip} & 2023 & Sticker sentiment. & \xmark  & \xmark & \xmark \\
 & SER30K \cite{SER30K} & 2022 & Sticker sentiment. & \cmark & \xmark & \xmark \\ \midrule

\textit{Captions} & TGIF \cite{tgif} & 2016 & GIF captioning. & \cmark & \xmark & \xmark \\
 & StickerTag \cite{stickertag} & 2024 & Sticker-tag. & \xmark & NA & NA \\ \midrule

\textit{Multimedia} & VSD2M \cite{vsd2m} & 2025 & Animated sticker dataset. & \cmark & \xmark & \xmark \\ 
 & ChineseB2B \cite{chinesebqb} & 2000 & Sticker images. & \cmark & \xmark & \xmark \\ \midrule

\textit{Retrieval} & PerSRV \cite{chee2024persrv} & 2024 & Sticker-query pairs. & \xmark & ZH & \xmark \\
 & \textbf{StickerQueries (Ours)} & \textbf{2025} & \textbf{Sticker-query pairs.} & \textbf{\cmark} & \textbf{EN, ZH} & \textbf{\cmark} \\

\bottomrule
\end{tabular}
\label{tab:query_dataset_comparison}
\end{table*}

\subsection{Sticker Semantic Datasets}
With the onset of instant messaging, we have observed significant advancements in sticker dataset retrieval. However, given the complexity and diversity of sticker tasks, existing datasets still represent only the tip of the iceberg. We broadly categorize these datasets into five groups: dialogues, sentiment, captions, multimedia and sticker queries.

Dialogue datasets typically consist of truncated conversations containing stickers \cite{learning-to-respond-2021, learning-to-respond-with-stickers-2020, MOD, stickerint, mcdscs, stickerconv, multichat}. In contrast, U-Sticker \cite{chee2025106kmultitopicmultilingualconversational} uniquely provides continuous user information, temporal context, and domain-specific sticker conversations. While some of these datasets are publicly available, others remain private or inaccessible. Sentiment datasets focus on predicting the sentiment conveyed by stickers, either as binary positive/negative labels \cite{CSMSA} or through more fine-grained classifications \cite{SER30K}. Certain datasets, such as Sticker820K \cite{stickerclip}, also include supplementary information like OCR text and descriptive metadata. Within this category, SER30K \cite{SER30K} is publicly available, whereas others are not. Caption datasets contain stickers with textual elements. For instance, TGIF \cite{tgif} provides captions for GIFs. Unfortunately, StickerTag \cite{stickertag} is unavailable, preventing further analysis. Multimedia datasets include raw sticker images for general use or generation tasks. ChineseB2B \cite{chinesebqb} offers a large collection of sticker images and VSD2M \cite{vsd2m} contains an extensive repository of sticker animations.

Finally, retrieval datasets pair user search queries with corresponding sticker images. These datasets are critical for capturing the underlying intent behind sticker usage through concise, high-precision keywords. Unfortunately, the only other sticker–query pair dataset, PerSRV \cite{chee2024persrv}, is not publicly available for analysis. Although PerSRV provides an open-source model, its training is limited to Chinese queries, restricting its applicability to English-language research. Furthermore, cultural differences mean that naive translation between languages often fails to convey the true intent of sticker usage. Our StickerQueries dataset addresses these limitations by offering bilingual (English and Chinese) queries crafted by native speakers, thereby bridging semantic understanding across cultural contexts and enabling more effective retrieval and recommendation systems.

\subsection{Image Captioning}
Recent models like BLIP and BLIP-2 \cite{li2022blip, li2023blip2}, as well as GIT \cite{wang2022git} and OFASys \cite{wang2022ofa}, have advanced image captioning through large-scale vision-language pretraining and unified generative modeling. VinVL \cite{zhang2021vinvl} and PaLI \cite{chen2023pali} further improve visual representations and multilingual capabilities. While these models produce accurate and fluent descriptions, they focus mainly on tangible visual semantics. In contrast, sticker search queries mainly revolves around emotional expression, which traditional captioning models are not designed to capture, making them inadequate for emotion-driven query generation.

\subsection{Large Language Model for Sticker Semantic Understanding}
Large Language Models (LLMs), such as ChatGPT, LLaMA, and their multimodal extensions \cite{openai2024gpt4technicalreport, chang2024survey, touvron2023llamaopenefficientfoundation, flamingo, li2022blip, li2023blip, zhu2023minigpt, liu2023llava, instructblip, wang2023zero}, have shown strong capabilities in integrating visual and textual information. Models like Flamingo and InstructBLIP introduce cross-modal alignment mechanisms, while BLIP, MiniGPT, and LLaVA variants connect frozen vision encoders with LLMs to enable richer multimodal understanding. These models hold promise for sticker semantics, where emotional and contextual cues are key. However, vanilla LLMs still struggle to generate intent-aligned sticker queries. PerSRV \cite{chee2024persrv} demonstrates this gap, showing that supervised fine-tuning is needed but its robustness across languages remains largely questionable. This highlights the necessity for a sticker query dataset which can improve sticker understanding and retrieval tasks.

\subsection{Games with a Purpose (GWAP)}

GWAP leverages human engagement through gameplay \cite{vonahn2008designing} and the low-pressure environment makes it particularly effective in gathering emotionally rich or subjective data which is often difficult to obtain through traditional annotation pipelines. Methodologies include output-agreement games,  games, where players independently try to generate the same label for an input, encouraging consensus on common interpretations; inversion-problem games,, where one player describes an item and the other guesses it, capturing descriptive and emotionally resonant cues; and input-agreement games, where players judge whether they received the same input, supporting perceptual or semantic similarity annotations \cite{law2008inputagreement}. GWAPs are especially effective and enjoyable, especially when combined with engaging visual effects, to enhance data quality. This makes them ideal for the sticker annotation task.

\section{Motivation and Goals}
\subsection{Sticker Usage Behavior Survey}
To better understand user sticker usage behavior, we conducted a survey covering variables such as age, primary language, average number of text messages sent before a sticker is used, the number of stickers used per day, and the motivations behind sticker usage.

We collected a total of \textbf{192} responses from a linguistically diverse user base and present our findings here. Interestingly, both genders reported the same average daily sticker usage—approximately \textbf{7.55 stickers per day}. While younger users generally tend to send more stickers, users across all age groups expressed genuine enjoyment in using them. On average, participants reported \textbf{sending a sticker every four text messages}. Notably, \textbf{none of the 192 respondents reported zero sticker usage}, underscoring the ubiquity of stickers in digital communication.

\begin{figure}[h]
    \centering
    \includegraphics[width=0.8\linewidth]{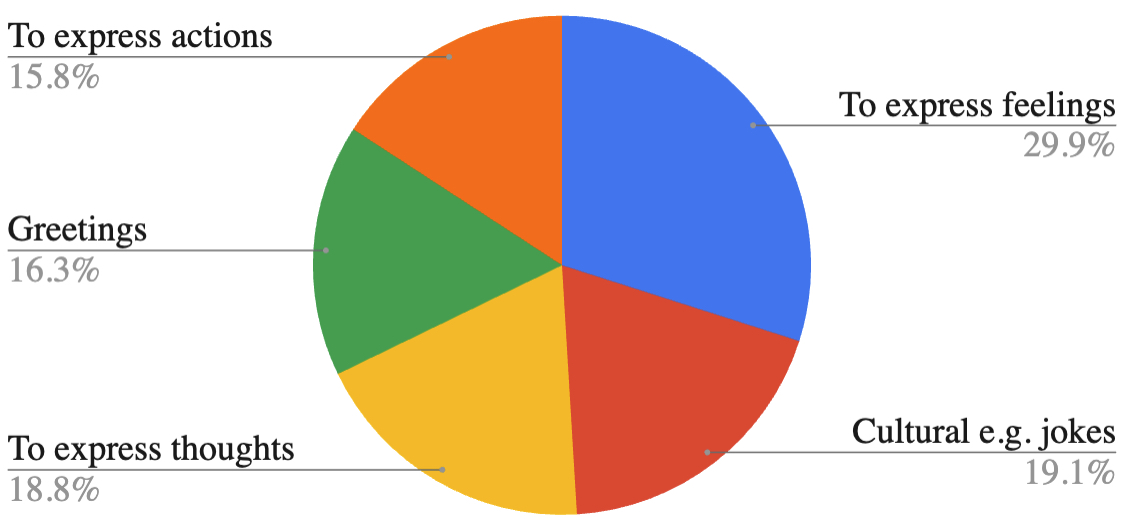}
    \caption{Emotional expression emerged as the most common reason for using stickers, while other motivations—such as actions, greetings and thoughts—were mentioned with relatively equal frequency.}
    \label{fig:sticker-usage-purpose}
\end{figure}

As shown in Fig~\ref{fig:sticker-usage-purpose}, the most common reason for using stickers is emotional expression, such as conveying feelings like happiness, sadness, or confusion. While this was the most dominant purpose cited by participants, other motivations were mentioned with relatively equal frequency. Users also mentioned using stickers for politeness or to clarify tone. These diverse uses reflect the rich communicative potential of stickers, not only do stickers enhance conversations but also make digital interactions more expressive, nuanced, and enjoyable. In light of these insights, it becomes clear that stickers play a pivotal role in digital communication—used frequently, across age groups, and for a wide range of expressive functions. The universality and versatility of sticker usage reveal a communication tool that is both culturally rich and deeply embedded in everyday interaction. With a vast and engaged user base, and an array of meaningful use cases, further exploration into sticker behavior holds tremendous potential—not only to advance human-computer interaction but to enrich how we understand and design for user expression. 



\subsection{Dataset Goals}
With this motivation in mind, we intend to develop the first multilingual sticker queries dataset publicably available. We establish the following goals.

\begin{itemize}
    \item \textbf{FAIR Compliance} \cite{fair_principles}: Ensure the dataset is openly accessible, well-documented, and licensed for reproducibility and reuse.
    \item \textbf{Natural Associations}: Focus on commonly used stickers and intuitive textual expressions, avoiding overly niche or esoteric cases.
    \item \textbf{Multilingual Coverage}: Avoid naive translations by incorporating various languages.
    \item \textbf{High Quality Representative Queries}: Capture expressions used by native speakers, including local slang and region-specific terms.
    \item \textbf{Diverse Expressions}: Reflect varied phrasing for similar intents (e.g., “haha,” “laugh out loud").
    \item \textbf{Resonant Queries}: Ensure queries are broadly understandable and relatable, not limited to a narrow subgroup.
\end{itemize}

\section{Sticktionary: Game-Based Sticker Semantics Annotation Framework}

\subsection{Game Design}
\begin{figure*}
    \centering
    \includegraphics[width=0.88\linewidth]
    {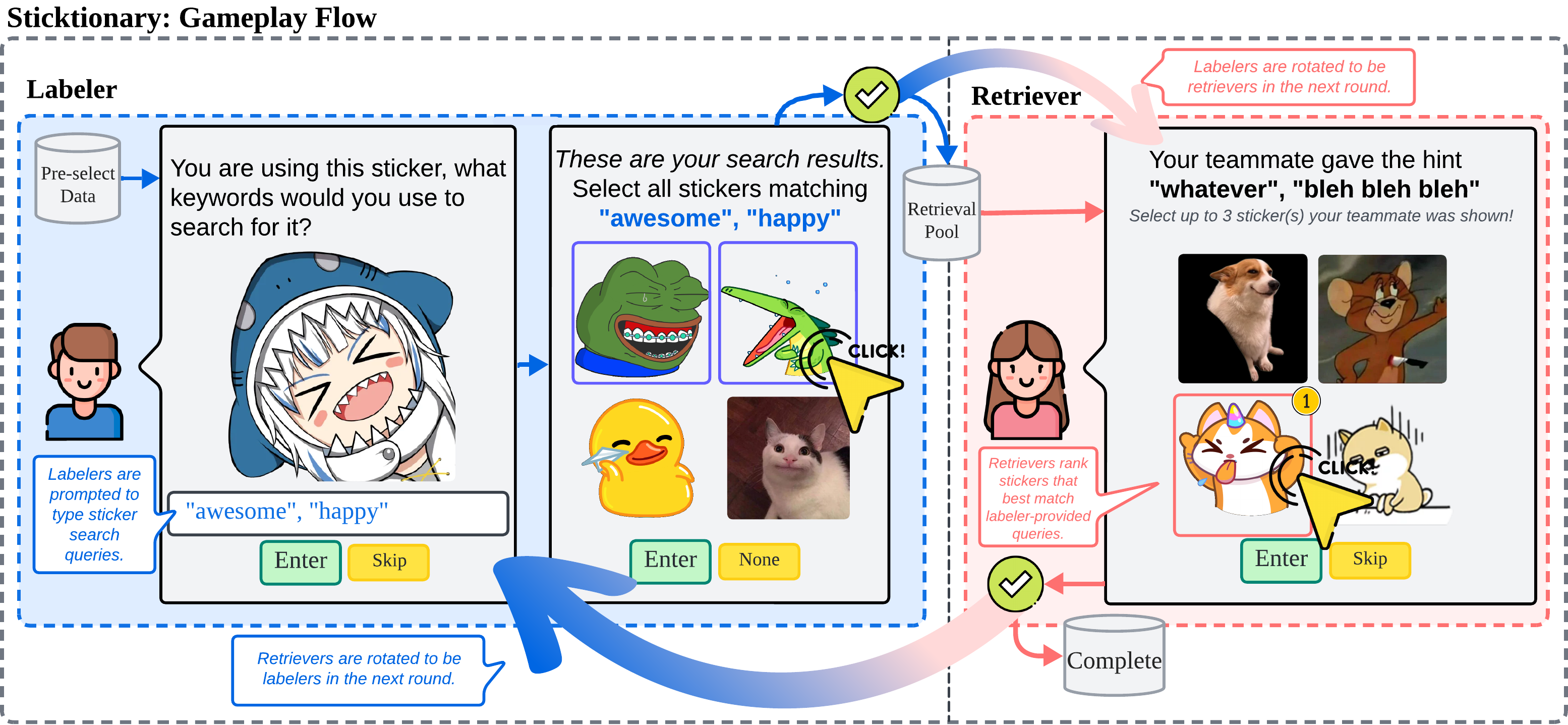}

    \caption{Sticktionary: A game-based framework for sticker semantics annotation. Players are randomly assigned initial roles and alternate between labeling and retrieving—annotating stickers with textual hints and identifying stickers based on those hints. Tasks are drawn from a curated pool and may be skipped. Annotated stickers enter a temporary retrieval pool, and successful retrievals mark task completion.}
    \label{fig:gameplay}
\end{figure*}


Notably, collecting high-quality representative queries, diverse expressions, and broadly resonant phrases is particularly challenging in traditional annotation pipelines. Agreeableness is especially difficult to model since sticker search is subjective, and ensuring high-quality representation often demands significant annotator effort and attention, potentially prolonging the annotation process. To address these challenges, we leverage the engaging and expressive visual modality of stickers and propose a game-based annotation framework, \textbf{Sticktionary}, illustrated in Fig~\ref{fig:gameplay}, is inspired by the GWAP paradigm. Sticktionary features two alternating roles—\textit{labeler} and \textit{retriever}—designed to collect and evaluate sticker search queries through interactive and enjoyable gameplay.

\textit{Labeler}
In each round, the labeler is shown a sticker and asked imagine themselves searching for the sticker, in which they are expected to key in search queries. Then, their queries are used to search and a preliminary sticker result is shown. This imaginative state prevents player from merely describing the features of the sticker. Based on this, they may revise their queries accordingly. Players are also instructed to avoid repetitive search queries, and purely visual descriptions. 

\textit{Retriever}
The retriever then receives the labeler's queries and must identify the correct sticker from a set of candidates, selecting and ranking up to three. Additionally, retrievers may optionally contribute new query suggestions for any sticker they feel confident about, encouraging the organic collection of emotionally resonant or imaginative queries. The outcome of the retrieval (successful or not) is shared with the labeler as feedback to refine future queries. Should the players feel that either task is too challenging they can choose to skip. 


\subsection{Game Settings}
\textit{Player Recruitment.}
We recruit participants by incentivizing top performers with themed prizes featuring the university's mascot panda, along with a participation reward of 30 RMB for completing an observable annotation session.

\textit{Data Selection.}
To ensure the goal of natural sticker associations, we select messages preceded by at least twenty words of dialogue, filtering out command-driven interactions and exchanges dominated by brief replies. This ensures that the stickers appear in expressive, emotionally grounded contexts. We apply these filters to the U-Sticker dataset \cite{chee2025106kmultitopicmultilingualconversational}, followed by a manual review. The final collection consists of 1,136 non-duplicated conversations, each curated for contextual richness and intuitive alignment.

\subsection{Goal-orientated}
\textit{Ensuring High Quality Queries.}
Sticktionary’s alternating roles sustain attention and reduce fatigue, boosting annotation quality. Points reward precise queries that retrieve the correct sticker, naturally validating their effectiveness. This retrieval-based gameplay also provides a natural validation signal, as successful retrievals indicate effective and contextually appropriate queries.


\textit{Ensuring Resonance.}
 Multiple successful retrieval from also indicates shared understanding and a resonant query. In addition, real-time feedback loops nudge players toward phrasing that feels universally intuitive, rather than overly niche or idiosyncratic.

\textit{Ensuring Diversity.}
Lastly, the involvement of more than one round of annotation can ensure diversity and variety. Each participant brings their own linguistic and cultural lens, contributing to a rich set of expressions for the same sticker. Furthermore, the reliance on retrievability as a success criterion naturally filters out repetitive or low-impact queries, ensuring that only distinct and effective expressions are retained.

\textit{Ensuring Multilingual Coverage, Representative Queries and Natural Associations.}
Players participate exclusively in their native language, ensuring authentic and culturally grounded expressions. Our data filtering process ensures that the stickers appear in expressive, emotionally grounded contexts.


These mechanisms work together—role-switching, feedback, and incentives—to ensure annotations are high-quality, diverse, and broadly resonant.

\subsection{Game Statistics}
\begin{table}[htbp]
\caption{Summary of player participation and task distribution across English and Chinese versions of the game after outlier removal. The dataset reflects contributions from over 60 players and approximately 60 hours of gameplay.}
\centering
\begin{tabular}{lrr}
\toprule
\textbf{Metric} & \textbf{English} & \textbf{Chinese} \\
\midrule
\# Label-role players       & 42    & 18    \\
\# Retrieval-role players     & 43    & 17    \\
Avg. tasks per label user & 61.1  & 78.7  \\
Avg. tasks per retrieve user & 55.0  & 68.5  \\
\bottomrule
\end{tabular}
\label{tab:players}
\end{table}

In total, we observed 60 hours worth of gameplay sessions which involved over 40 English players and 22 Chinese players. Table~\ref{tab:players} provides a comprehensive snapshot of participant engagement and workload distribution across labeling and retrieval activities. 

For successfully retrieved samples, we append them directly. Failures undergo manual review, where most ill retrievals stem from a flood of semantically similar stickers—making retrieval results actually reasonable and acceptable. Inaccurate queries are discarded. Lastly, obvious spelling errors are corrected.

\section{StickerQueries}
\subsection{Statistics}
\begin{table}[htbp]
\caption{StickerQueries comprises over 1,700 multilingual sticker-query pairs, each annotated by at least two annotators, offering a rich, diverse set of natural expressions.}
\centering
\begin{tabular}{lrr}
\toprule
\textbf{Metric} & \textbf{English} & \textbf{Chinese} \\
\midrule
Unique sticker query pair        & 1,116    & 615    \\
Unique terms     & 5,347    & 1,944    \\
Avg. \# queries per sticker & 6.77  & 4.60  \\
\bottomrule
\end{tabular}
\label{tab:overall_stats}
\end{table}
Table \ref{tab:overall_stats} presents overview of the StickerQueries dataset statistics, highlighting its scale, linguistic diversity, and annotation coverage. For the following steps, we use Jieba for Chinese tokenization~\cite{jieba}.

\begin{figure}[htbp]
    \centering
    \includegraphics[width=0.78\linewidth]{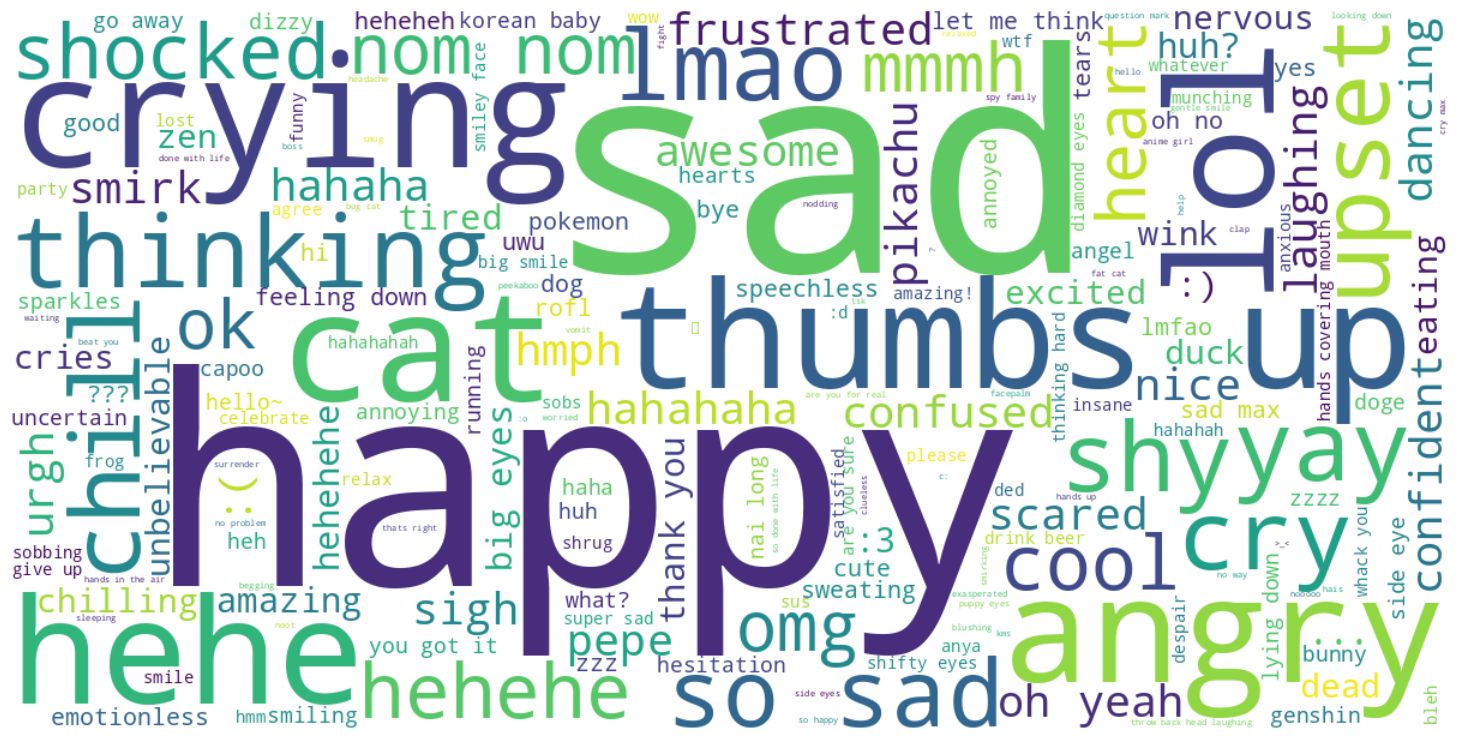}
    \includegraphics[width=0.78\linewidth]{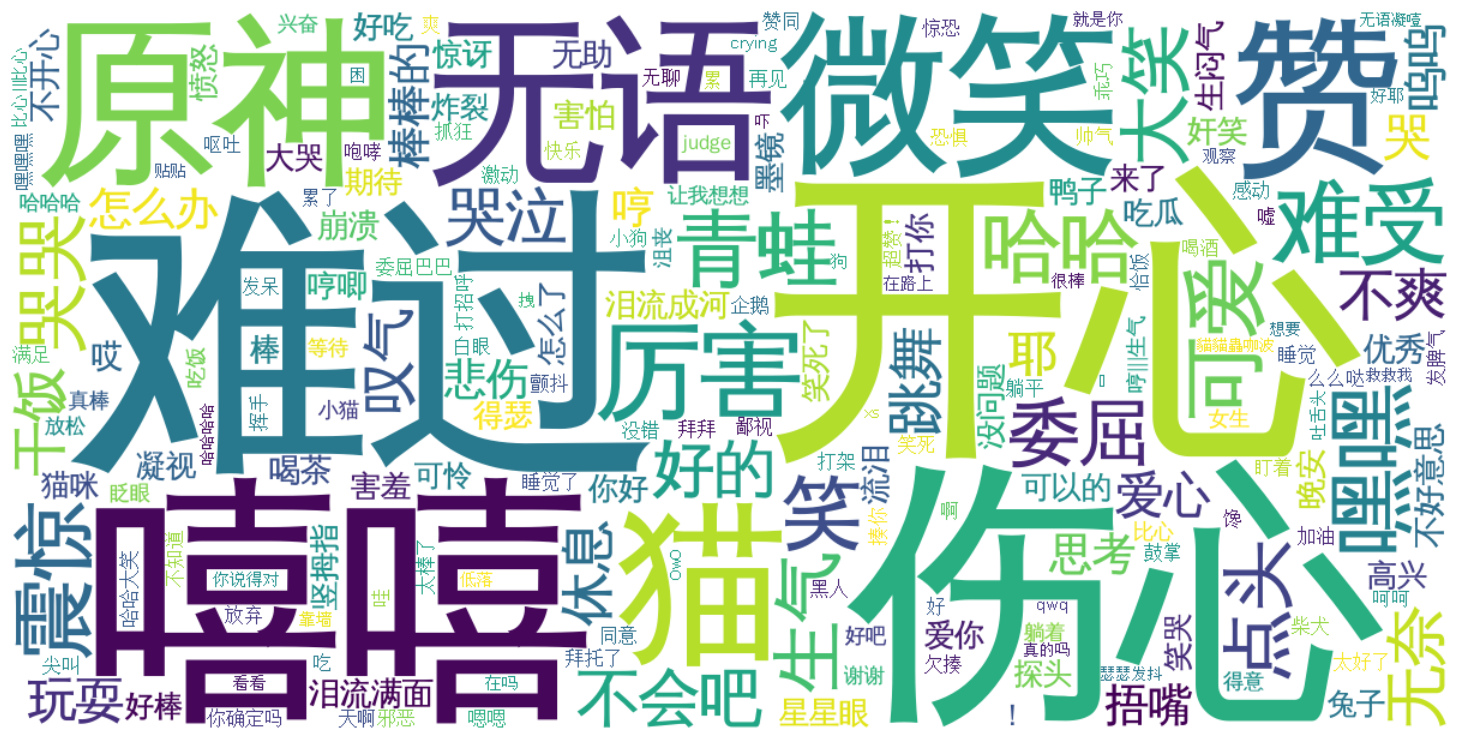}
    \caption{Word cloud of the English (top) and Chinese (bottom) StickerQueries, feelings dominate the visual.}
    \label{fig:english_wordcloud}
\end{figure}


\begin{figure}[htbp]
    \centering
    \includegraphics[width=0.87\linewidth]{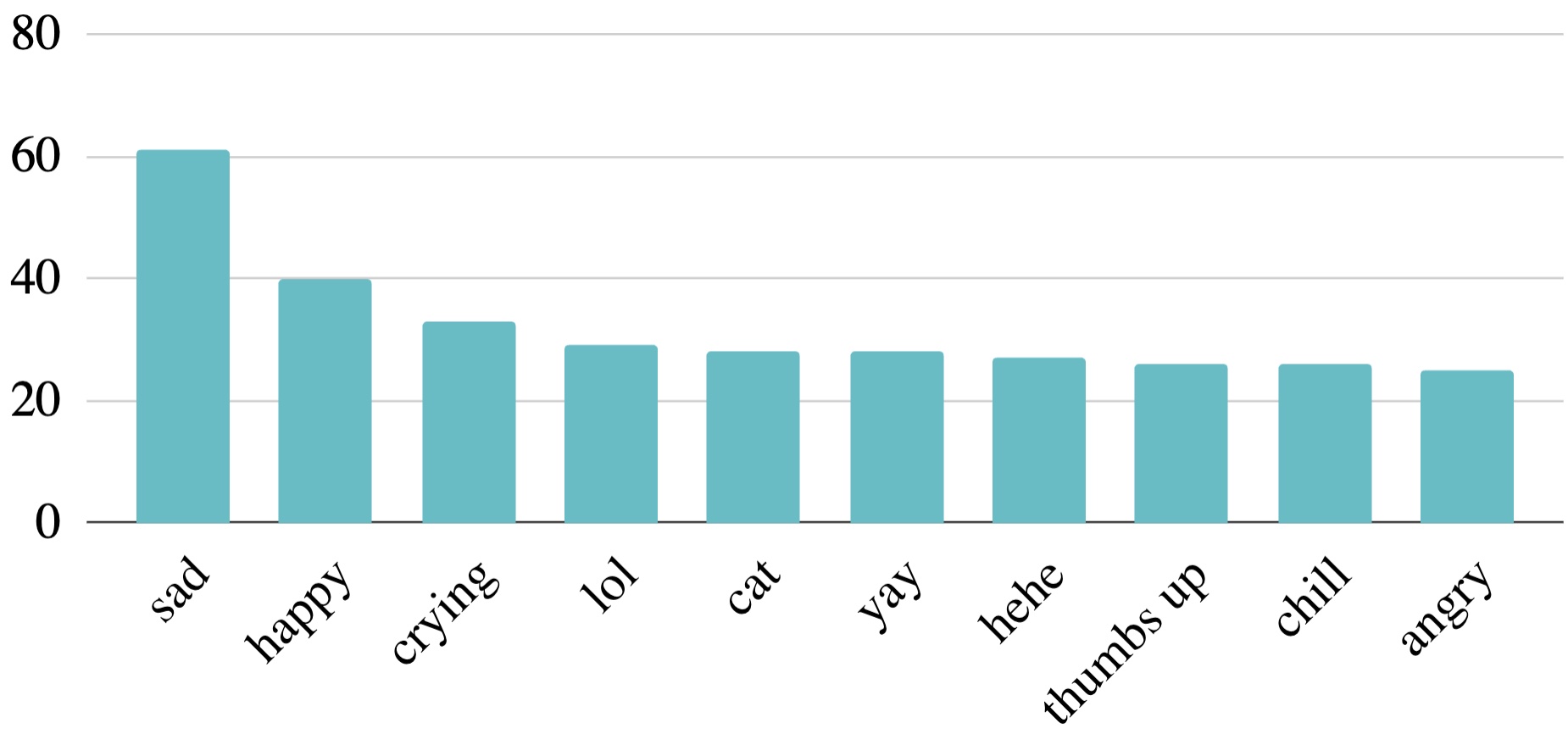}
    \includegraphics[width=0.87\linewidth]{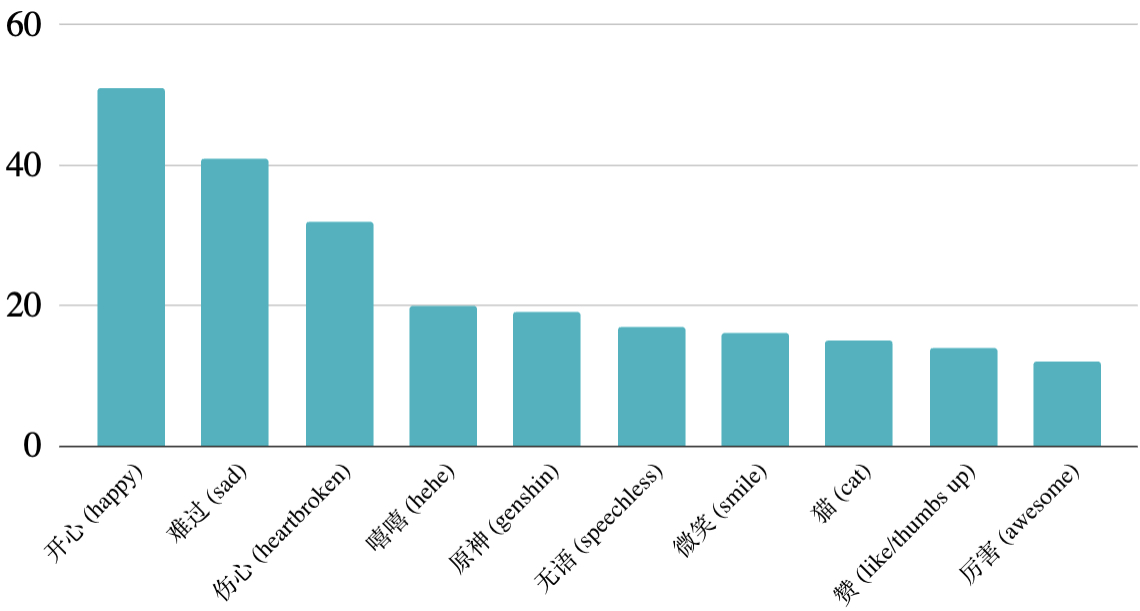}
    \caption{Frequency distribution of the top 10 most common English (top) and Chinese (bottom) query words in the StickerQueries dataset.}
    \label{fig:english_top_words}
\end{figure}

From Figures~\ref{fig:english_top_words} and \ref{fig:english_wordcloud}, we observe that emotional expressions dominate sticker queries: 9 of the top 10 terms in both English and Chinese relate to feelings rather than actions or visuals. This supports the view that users are often guided by internal emotional states when searching, rather than external depictions. Although survey participants emphasized action-oriented queries (Fig ~\ref{fig:sticker-usage-purpose}, annotation patterns preliminarily reveals a stronger tendency toward emotional resonance—affirming stickers’ role as an affective communication medium.

Additionally, while English and Chinese users share many emotional sentiments, subtle cross-linguistic and cultural differences emerge. English-exclusive terms include lol, chill, and zen, whereas culturally rooted Chinese expressions like eating melon (watching gossip), do rice (eating a meal), and awesome awesome one (awesome) appear. These highlight the limitations of naive translation and emphasize the cultural richness captured in the StickerQueries dataset




\subsection{Interannotator Similarity}

\begin{table}[htbp]
\centering
\caption{Multilingual sticker query similarity evaluation. BLEU and ROUGE (R) are low due to short queries, while BERT-based (B) and cosine similarity (CS) show higher semantic alignment.}
\label{tab:multilingual_metrics}
\begin{tabular}{ccccccc}
\toprule 
& \textbf{BLEU} & \textbf{R-L} & \textbf{CS} & \textbf{B (P)} & \textbf{B (R)} & \textbf{B (F1)} \\
\midrule
 EN & 0.0756 & 0.2796 & 0.7131 & 0.6710 & 0.6666 & 0.6628 \\ \midrule
 ZH & 0.0132 & 0.0195 & 0.7728 & 0.7061 & 0.6983 & 0.6986 \\
\bottomrule
\end{tabular}
\end{table}

We use the pretrained BERT-base models \cite{devlin2019bert} released by Google for English and Chinese to compute both cosine similarity and BERTScore \cite{zhang2019bertscore} between query annotations. ROUGE-L \cite{lin2004rouge} are also reported for comparison with traditional string-based metrics.



English annotations show higher lexical overlap, with BLEU (0.0756) and ROUGE scores (e.g., ROUGE-L: 0.2796), indicating strong n-gram similarity. Semantic metrics also confirm good meaning alignment (cosine: 0.7131, BERT F1: 0.6628). Chinese annotations score lower on lexical metrics (BLEU: 0.0132, ROUGE-L: 0.0195), likely due to greater paraphrasing. However, they achieve higher semantic similarity (cosine: 0.7728, BERT F1: 0.6986), indicating strong meaning preservation despite fewer surface overlaps. Following prior work \cite{zhang2019bertscore}, a BERTScore F1 above 0.65 indicates moderate semantic agreement in open-ended generation tasks. Our dataset achieves 0.66–0.70, validating the reliability of annotations while retaining expressive diversity.

\section{Applications}
\subsection{Query Generation}

We split the data into 80\% training and 20\% validation and test sets, then fine-tuned LLaVA-1.5B-7B on our StickerQueries dataset. 
Fine-tuning on StickerQueries dataset leads to clear improvements across all evaluation metrics for both English and Chinese, as shown in Table ~\ref{tab:llava_performance}. This demonstrates the dataset’s effectiveness in enhancing the model’s ability to generate accurate, contextually relevant sticker queries.

\begin{table}[htbp]
\centering
\caption{Performance of LLaVA-1.5-7B on StickerQueries, showing zero-shot vs fine-tuned results. Significant improvements demonstrate effective adaptation to both English and Chinese query generation. R short for ROUGE.}
\label{tab:llava_performance}
\begin{tabular}{llcccc}
\toprule
 & \textbf{Model} & \textbf{BLEU} & \textbf{R-2} & \textbf{R-L} & \textbf{CS} \\
\midrule
EN & Vanilla VLM & 11.08 & 2.82 & 19.22 & 0.78 \\
& SFT VLM & 15.62 & 8.26 & 27.05 & 0.81 \\ \midrule

\addlinespace[0.5em]
ZH & Vanilla VLM & 0.51 & 0.03 & 2.26 & 0.63 \\
& SFT VLM & 6.34 & 2.97 & 24.26 & 0.80 \\
\bottomrule
\end{tabular}
\end{table}

\subsection{Sticker Retrieval}

\begin{table}[htbp]
\centering
\caption{Recall@K performance comparison across query generation methods.}
\label{tab:performance}
\begin{tabular}{@{}lcccc@{}}
\toprule
\multirow{2}{*}{Model} & \multicolumn{2}{c}{English} & \multicolumn{2}{c}{Chinese} \\
\cmidrule(lr){2-3} \cmidrule(lr){4-5}
 & R@1 & R@50 & R@1 & R@50 \\
\midrule
Context & 0.018 & 0.297 & 0.008 & 0.407 \\ 
BLIP2 & 0.032 & 0.248 & 0.005 & 0.225 \\
Vanilla VLM & 0.153 & 0.644 & 0.016 & 0.423 \\
\midrule
\textbf{SFT VLM} & \textbf{0.180} & \textbf{0.680} & \textbf{0.041} & \textbf{0.455} \\
\bottomrule
\end{tabular}
\end{table}




On the same 20\% test split, we compare three sticker semantic baseline methods for the sticker retrieval task to generate queries.
\begin{itemize}
    \item \textbf{Context:} Take previous utterance as query.
    \item \textbf{BLIP2} \cite{li2023blip2}: Generate queries via image captioning.
    \item \textbf{Vanilla VLM} \cite{liu2023llava}: Zero-shot LLaVA-1.5B-7B.
    \item \textbf{SFT VLM (Ours):} Finetuned LLaVA-1.5B-7B model.
\end{itemize}
We generate test candidate pools using the above baselines and use BM25 \cite{robertson2009probabilistic} for retrieval. 
The fine-tuned LLaVA model trained on the StickerQueries dataset achieves the highest Recall@K scores at every cutoff, indicating it generates more accurate and relevant queries for sticker retrieval. Compared to zero-shot baselines like Vanilla VLM, BLIP2 image captions, and direct conversation context, our model provides a significantly stronger candidate pool for BM25 retrieval, underscoring the benefits of domain-specific fine-tuning.


\section{Limitations and Future Work}
We acknowledge our dataset’s limitations, including a primarily young adult annotation pool and a relatively small size. Expanding it with more survey respondents, diverse participants, additional annotations, and multilingual data would likely improve performance, robustness, and understanding of sticker usage behavior. 

\section{Conclusion}
In this work, we propose StickerQueries, a multilingual sticker semantics understanding dataset for sticker understanding and retrieval, built alongside Sticktionary, a gamified annotation framework. Together, they capture rich, high-quality queries and culturally grounded sticker semantics. Fine-tuning vision-language models on our dataset improves query generation and retrieval, highlighting the limits of general-purpose models and the need for specialized data in high-context multimodal tasks. These resources provide a strong foundation for future work in personalized visual communication, semantic understanding, and cross-lingual sticker interaction.
\clearpage

\bibliographystyle{ACM-Reference-Format}
\balance
\bibliography{sample-base}


\begin{thebibliography}{40}


\ifx \showCODEN    \undefined \def \showCODEN     #1{\unskip}     \fi
\ifx \showDOI      \undefined \def \showDOI       #1{#1}\fi
\ifx \showISBNx    \undefined \def \showISBNx     #1{\unskip}     \fi
\ifx \showISBNxiii \undefined \def \showISBNxiii  #1{\unskip}     \fi
\ifx \showISSN     \undefined \def \showISSN      #1{\unskip}     \fi
\ifx \showLCCN     \undefined \def \showLCCN      #1{\unskip}     \fi
\ifx \shownote     \undefined \def \shownote      #1{#1}          \fi
\ifx \showarticletitle \undefined \def \showarticletitle #1{#1}   \fi
\ifx \showURL      \undefined \def \showURL       {\relax}        \fi
\providecommand\bibfield[2]{#2}
\providecommand\bibinfo[2]{#2}
\providecommand\natexlab[1]{#1}
\providecommand\showeprint[2][]{arXiv:#2}

\bibitem[Achiam et~al\mbox{.}(2024)]%
        {openai2024gpt4technicalreport}
\bibfield{author}{\bibinfo{person}{Josh Achiam}, \bibinfo{person}{Steven Adler}, \bibinfo{person}{Sandhini Agarwal}, {and} \bibinfo{person}{Others}.} \bibinfo{year}{2024}\natexlab{}.
\newblock \bibinfo{title}{GPT-4 Technical Report}.
\newblock
\newblock
\showeprint[arxiv]{2303.08774}~[cs.CL]
\urldef\tempurl%
\url{https://arxiv.org/abs/2303.08774}
\showURL{%
\tempurl}


\bibitem[Alayrac et~al\mbox{.}(2022)]%
        {flamingo}
\bibfield{author}{\bibinfo{person}{Jean-Baptiste Alayrac}, \bibinfo{person}{Jeff Donahue}, \bibinfo{person}{Pauline Luc}, \bibinfo{person}{Antoine Miech}, \bibinfo{person}{Iain Barr}, \bibinfo{person}{Yana Hasson}, \bibinfo{person}{Karel Lenc}, \bibinfo{person}{Arthur Mensch}, \bibinfo{person}{Katherine Millican}, \bibinfo{person}{Malcolm Reynolds}, \bibinfo{person}{Roman Ring}, \bibinfo{person}{Eliza Rutherford}, \bibinfo{person}{Serkan Cabi}, \bibinfo{person}{Tengda Han}, \bibinfo{person}{Zhitao Gong}, \bibinfo{person}{Sina Samangooei}, \bibinfo{person}{Marianne Monteiro}, \bibinfo{person}{Jacob~L Menick}, \bibinfo{person}{Sebastian Borgeaud}, \bibinfo{person}{Andy Brock}, \bibinfo{person}{Aida Nematzadeh}, \bibinfo{person}{Sahand Sharifzadeh}, \bibinfo{person}{Miko\l~aj Bi\'{n}kowski}, \bibinfo{person}{Ricardo Barreira}, \bibinfo{person}{Oriol Vinyals}, \bibinfo{person}{Andrew Zisserman}, {and} \bibinfo{person}{Kar\'{e}n Simonyan}.} \bibinfo{year}{2022}\natexlab{}.
\newblock \showarticletitle{Flamingo: a Visual Language Model for Few-Shot Learning}. In \bibinfo{booktitle}{\emph{Advances in Neural Information Processing Systems}}, \bibfield{editor}{\bibinfo{person}{S.~Koyejo}, \bibinfo{person}{S.~Mohamed}, \bibinfo{person}{A.~Agarwal}, \bibinfo{person}{D.~Belgrave}, \bibinfo{person}{K.~Cho}, {and} \bibinfo{person}{A.~Oh}} (Eds.), Vol.~\bibinfo{volume}{35}. \bibinfo{publisher}{Curran Associates, Inc.}, \bibinfo{pages}{23716--23736}.
\newblock
\urldef\tempurl%
\url{https://proceedings.neurips.cc/paper_files/paper/2022/file/960a172bc7fbf0177ccccbb411a7d800-Paper-Conference.pdf}
\showURL{%
\tempurl}


\bibitem[{Association for Computational Linguistics 2024} and Zhang(2024)]%
        {stickerconv}
\bibfield{author}{\bibinfo{person}{{Association for Computational Linguistics 2024}} {and} \bibinfo{person}{Yiqun Zhang}.} \bibinfo{year}{2024}\natexlab{}.
\newblock \bibinfo{title}{STICKERCONV: Generating Multimodal Empathetic Responses from Scratch}.
\newblock
\newblock
\urldef\tempurl%
\url{https://doi.org/10.48448/6S4N-D973}
\showDOI{\tempurl}


\bibitem[Chang et~al\mbox{.}(2024)]%
        {chang2024survey}
\bibfield{author}{\bibinfo{person}{Yupeng Chang}, \bibinfo{person}{Xu Wang}, \bibinfo{person}{Jindong Wang}, \bibinfo{person}{Yuan Wu}, \bibinfo{person}{Linyi Yang}, \bibinfo{person}{Kaijie Zhu}, \bibinfo{person}{Hao Chen}, \bibinfo{person}{Xiaoyuan Yi}, \bibinfo{person}{Cunxiang Wang}, \bibinfo{person}{Yidong Wang}, {et~al\mbox{.}}} \bibinfo{year}{2024}\natexlab{}.
\newblock \showarticletitle{A survey on evaluation of large language models}.
\newblock \bibinfo{journal}{\emph{ACM Transactions on Intelligent Systems and Technology}} \bibinfo{volume}{15}, \bibinfo{number}{3} (\bibinfo{year}{2024}), \bibinfo{pages}{1--45}.
\newblock


\bibitem[Chee et~al\mbox{.}(2025)]%
        {chee2025106kmultitopicmultilingualconversational}
\bibfield{author}{\bibinfo{person}{Heng Er~Metilda Chee}, \bibinfo{person}{Jiayin Wang}, \bibinfo{person}{Zhiqiang Guo}, \bibinfo{person}{Weizhi Ma}, \bibinfo{person}{Qinglang Guo}, {and} \bibinfo{person}{Min Zhang}.} \bibinfo{year}{2025}\natexlab{}.
\newblock \bibinfo{title}{A 106K Multi-Topic Multilingual Conversational User Dataset with Emoticons}.
\newblock
\newblock
\showeprint[arxiv]{2502.19108}~[cs.IR]
\urldef\tempurl%
\url{https://arxiv.org/abs/2502.19108}
\showURL{%
\tempurl}


\bibitem[Chee et~al\mbox{.}(2024)]%
        {chee2024persrv}
\bibfield{author}{\bibinfo{person}{Heng Er~Metilda Chee}, \bibinfo{person}{Jiayin Wang}, \bibinfo{person}{Zhiqiang Guo}, \bibinfo{person}{Weizhi Ma}, {and} \bibinfo{person}{Min Zhang}.} \bibinfo{year}{2024}\natexlab{}.
\newblock \showarticletitle{PerSRV: Personalized Sticker Retrieval with Vision-Language Model}.
\newblock \bibinfo{journal}{\emph{arXiv preprint arXiv:2410.21801}} (\bibinfo{year}{2024}).
\newblock


\bibitem[Chen et~al\mbox{.}(2023)]%
        {chen2023pali}
\bibfield{author}{\bibinfo{person}{Yen-Chun Chen} {et~al\mbox{.}}} \bibinfo{year}{2023}\natexlab{}.
\newblock \showarticletitle{PaLI: A Jointly-Scaled Multilingual Language–Image Model}. In \bibinfo{booktitle}{\emph{CVPR}}.
\newblock


\bibitem[Dai et~al\mbox{.}(2024)]%
        {instructblip}
\bibfield{author}{\bibinfo{person}{Wenliang Dai}, \bibinfo{person}{Junnan Li}, \bibinfo{person}{Dongxu Li}, \bibinfo{person}{Anthony Meng~Huat Tiong}, \bibinfo{person}{Junqi Zhao}, \bibinfo{person}{Weisheng Wang}, \bibinfo{person}{Boyang Li}, \bibinfo{person}{Pascale Fung}, {and} \bibinfo{person}{Steven Hoi}.} \bibinfo{year}{2024}\natexlab{}.
\newblock \showarticletitle{InstructBLIP: towards general-purpose vision-language models with instruction tuning}. In \bibinfo{booktitle}{\emph{Proceedings of the 37th International Conference on Neural Information Processing Systems}} (New Orleans, LA, USA) \emph{(\bibinfo{series}{NIPS '23})}. \bibinfo{publisher}{Curran Associates Inc.}, \bibinfo{address}{Red Hook, NY, USA}, Article \bibinfo{articleno}{2142}, \bibinfo{numpages}{18}~pages.
\newblock


\bibitem[Devlin et~al\mbox{.}(2019)]%
        {devlin2019bert}
\bibfield{author}{\bibinfo{person}{Jacob Devlin}, \bibinfo{person}{Ming-Wei Chang}, \bibinfo{person}{Kenton Lee}, {and} \bibinfo{person}{Kristina Toutanova}.} \bibinfo{year}{2019}\natexlab{}.
\newblock \showarticletitle{BERT: Pre-training of Deep Bidirectional Transformers for Language Understanding}.
\newblock \bibinfo{journal}{\emph{arXiv preprint arXiv:1810.04805}} (\bibinfo{year}{2019}).
\newblock


\bibitem[Fei et~al\mbox{.}(2021)]%
        {MOD}
\bibfield{author}{\bibinfo{person}{Zhengcong Fei}, \bibinfo{person}{Zekang Li}, \bibinfo{person}{Jinchao Zhang}, \bibinfo{person}{Yang Feng}, {and} \bibinfo{person}{Jie Zhou}.} \bibinfo{year}{2021}\natexlab{}.
\newblock \showarticletitle{Towards expressive communication with internet memes: A new multimodal conversation dataset and benchmark}.
\newblock \bibinfo{journal}{\emph{arXiv preprint arXiv:2109.01839}} (\bibinfo{year}{2021}).
\newblock


\bibitem[Gao et~al\mbox{.}(2020)]%
        {learning-to-respond-with-stickers-2020}
\bibfield{author}{\bibinfo{person}{Shen Gao}, \bibinfo{person}{Xiuying Chen}, \bibinfo{person}{Chang Liu}, \bibinfo{person}{Li Liu}, \bibinfo{person}{Dongyan Zhao}, {and} \bibinfo{person}{Rui Yan}.} \bibinfo{year}{2020}\natexlab{}.
\newblock \showarticletitle{Learning to Respond with Stickers: A Framework of Unifying Multi-Modality in Multi-Turn Dialog}. In \bibinfo{booktitle}{\emph{Proceedings of The Web Conference 2020}} \emph{(\bibinfo{series}{WWW '20})}. \bibinfo{pages}{1138–1148}.
\newblock
\showISBNx{9781450370233}
\urldef\tempurl%
\url{https://doi.org/10.1145/3366423.3380191}
\showDOI{\tempurl}


\bibitem[Gao et~al\mbox{.}(2021)]%
        {learning-to-respond-2021}
\bibfield{author}{\bibinfo{person}{Shen Gao}, \bibinfo{person}{Xiuying Chen}, \bibinfo{person}{Li Liu}, \bibinfo{person}{Dongyan Zhao}, {and} \bibinfo{person}{Rui Yan}.} \bibinfo{year}{2021}\natexlab{}.
\newblock \showarticletitle{Learning to Respond with Your Favorite Stickers: A Framework of Unifying Multi-Modality and User Preference in Multi-Turn Dialog}.
\newblock \bibinfo{journal}{\emph{ACM Trans. Inf. Syst.}} \bibinfo{volume}{39}, \bibinfo{number}{2}, Article \bibinfo{articleno}{12} (\bibinfo{date}{Feb.} \bibinfo{year}{2021}), \bibinfo{numpages}{32}~pages.
\newblock
\showISSN{1046-8188}
\urldef\tempurl%
\url{https://doi.org/10.1145/3429980}
\showDOI{\tempurl}


\bibitem[Ge et~al\mbox{.}(2022)]%
        {CSMSA}
\bibfield{author}{\bibinfo{person}{Feng Ge}, \bibinfo{person}{Weizhao Li}, \bibinfo{person}{Haopeng Ren}, {and} \bibinfo{person}{Yi Cai}.} \bibinfo{year}{2022}\natexlab{}.
\newblock \showarticletitle{Towards exploiting sticker for multimodal sentiment analysis in social media: A new dataset and baseline}. In \bibinfo{booktitle}{\emph{Proceedings of the 29th International Conference on Computational Linguistics}}. \bibinfo{pages}{6795--6804}.
\newblock


\bibitem[Junyi(2012)]%
        {jieba}
\bibfield{author}{\bibinfo{person}{Sun Junyi}.} \bibinfo{year}{2012}\natexlab{}.
\newblock \bibinfo{title}{Jieba Chinese Text Segmentation: built to be the best Python Chinese word segmentation module.}
\newblock \bibinfo{howpublished}{\url{https://github.com/fxsjy/jieba}}.
\newblock
\newblock
\shownote{Accessed: 2025-05-30}.


\bibitem[Law and von Ahn(2009)]%
        {law2008inputagreement}
\bibfield{author}{\bibinfo{person}{Edith Law} {and} \bibinfo{person}{Luis von Ahn}.} \bibinfo{year}{2009}\natexlab{}.
\newblock \showarticletitle{Input-agreement: a new mechanism for collecting data using human computation games}. In \bibinfo{booktitle}{\emph{Proceedings of the SIGCHI Conference on Human Factors in Computing Systems}}. \bibinfo{pages}{1197--1206}.
\newblock


\bibitem[Li et~al\mbox{.}(2023a)]%
        {li2023blip2}
\bibfield{author}{\bibinfo{person}{Junnan Li} {et~al\mbox{.}}} \bibinfo{year}{2023}\natexlab{a}.
\newblock \showarticletitle{BLIP-2: Bootstrapping Language-Image Pre-training with Frozen Image Encoders and Large Language Models}. In \bibinfo{booktitle}{\emph{ICLR}}.
\newblock


\bibitem[Li et~al\mbox{.}(2023b)]%
        {li2023blip}
\bibfield{author}{\bibinfo{person}{Junnan Li}, \bibinfo{person}{Dongxu Li}, \bibinfo{person}{Silvio Savarese}, {and} \bibinfo{person}{Steven Hoi}.} \bibinfo{year}{2023}\natexlab{b}.
\newblock \showarticletitle{Blip-2: Bootstrapping language-image pre-training with frozen image encoders and large language models}. In \bibinfo{booktitle}{\emph{International conference on machine learning}}. PMLR, \bibinfo{pages}{19730--19742}.
\newblock


\bibitem[Li et~al\mbox{.}(2022)]%
        {li2022blip}
\bibfield{author}{\bibinfo{person}{Junnan Li}, \bibinfo{person}{Dongxu Li}, \bibinfo{person}{Caiming Xiong}, {and} \bibinfo{person}{Steven Hoi}.} \bibinfo{year}{2022}\natexlab{}.
\newblock \showarticletitle{Blip: Bootstrapping language-image pre-training for unified vision-language understanding and generation}. In \bibinfo{booktitle}{\emph{International conference on machine learning}}. PMLR, \bibinfo{pages}{12888--12900}.
\newblock


\bibitem[Li et~al\mbox{.}(2016)]%
        {tgif}
\bibfield{author}{\bibinfo{person}{Yuncheng Li}, \bibinfo{person}{Yale Song}, \bibinfo{person}{Liangliang Cao}, \bibinfo{person}{Joel Tetreault}, \bibinfo{person}{Larry Goldberg}, \bibinfo{person}{Alejandro Jaimes}, {and} \bibinfo{person}{Jiebo Luo}.} \bibinfo{year}{2016}\natexlab{}.
\newblock \bibinfo{title}{TGIF: A New Dataset and Benchmark on Animated GIF Description}.
\newblock
\newblock
\showeprint[arxiv]{1604.02748}~[cs.CV]
\urldef\tempurl%
\url{https://arxiv.org/abs/1604.02748}
\showURL{%
\tempurl}


\bibitem[Liang et~al\mbox{.}(2024)]%
        {stickerint}
\bibfield{author}{\bibinfo{person}{Bin Liang}, \bibinfo{person}{Bingbing Wang}, \bibinfo{person}{Zhixin Bai}, \bibinfo{person}{Qiwei Lang}, \bibinfo{person}{Mingwei Sun}, \bibinfo{person}{Kaiheng Hou}, \bibinfo{person}{Lanjun Zhou}, \bibinfo{person}{Ruifeng Xu}, {and} \bibinfo{person}{Kam-Fai Wong}.} \bibinfo{year}{2024}\natexlab{}.
\newblock \bibinfo{title}{Reply with Sticker: New Dataset and Model for Sticker Retrieval}.
\newblock
\newblock
\showeprint[arxiv]{2403.05427}~[cs.MM]
\urldef\tempurl%
\url{https://arxiv.org/abs/2403.05427}
\showURL{%
\tempurl}


\bibitem[Lin(2004)]%
        {lin2004rouge}
\bibfield{author}{\bibinfo{person}{Chin-Yew Lin}.} \bibinfo{year}{2004}\natexlab{}.
\newblock \showarticletitle{ROUGE: A package for automatic evaluation of summaries}. In \bibinfo{booktitle}{\emph{Text summarization branches out: Proceedings of the ACL-04 workshop}}. \bibinfo{pages}{74--81}.
\newblock


\bibitem[Liu et~al\mbox{.}(2023)]%
        {liu2023llava}
\bibfield{author}{\bibinfo{person}{Haotian Liu}, \bibinfo{person}{Chunyuan Li}, \bibinfo{person}{Qingyang Wu}, {and} \bibinfo{person}{Yong~Jae Lee}.} \bibinfo{year}{2023}\natexlab{}.
\newblock \showarticletitle{Visual Instruction Tuning}. In \bibinfo{booktitle}{\emph{NeurIPS}}.
\newblock


\bibitem[Liu et~al\mbox{.}(2022)]%
        {SER30K}
\bibfield{author}{\bibinfo{person}{Shengzhe Liu}, \bibinfo{person}{Xin Zhang}, {and} \bibinfo{person}{Jufeng Yang}.} \bibinfo{year}{2022}\natexlab{}.
\newblock \showarticletitle{SER30K: A large-scale dataset for sticker emotion recognition}. In \bibinfo{booktitle}{\emph{Proceedings of the 30th ACM International Conference on Multimedia}}. \bibinfo{pages}{33--41}.
\newblock


\bibitem[Robertson et~al\mbox{.}(2009)]%
        {robertson2009probabilistic}
\bibfield{author}{\bibinfo{person}{Stephen Robertson}, \bibinfo{person}{Hugo Zaragoza}, {et~al\mbox{.}}} \bibinfo{year}{2009}\natexlab{}.
\newblock \showarticletitle{The probabilistic relevance framework: BM25 and beyond}.
\newblock \bibinfo{journal}{\emph{Foundations and Trends{\textregistered} in Information Retrieval}} \bibinfo{volume}{3}, \bibinfo{number}{4} (\bibinfo{year}{2009}), \bibinfo{pages}{333--389}.
\newblock


\bibitem[Shi and Kong(2024)]%
        {mcdscs}
\bibfield{author}{\bibinfo{person}{Yuanchen Shi} {and} \bibinfo{person}{Fang Kong}.} \bibinfo{year}{2024}\natexlab{}.
\newblock \showarticletitle{Integrating Stickers into Multimodal Dialogue Summarization: A Novel Dataset and Approach for Enhancing Social Media Interaction}. In \bibinfo{booktitle}{\emph{Proceedings of the 32nd ACM International Conference on Multimedia}}. \bibinfo{pages}{9525--9534}.
\newblock


\bibitem[Touvron et~al\mbox{.}(2023)]%
        {touvron2023llamaopenefficientfoundation}
\bibfield{author}{\bibinfo{person}{Hugo Touvron}, \bibinfo{person}{Thibaut Lavril}, \bibinfo{person}{Gautier Izacard}, \bibinfo{person}{Xavier Martinet}, \bibinfo{person}{Marie-Anne Lachaux}, \bibinfo{person}{Timothée Lacroix}, \bibinfo{person}{Baptiste Rozière}, \bibinfo{person}{Naman Goyal}, \bibinfo{person}{Eric Hambro}, \bibinfo{person}{Faisal Azhar}, \bibinfo{person}{Aurelien Rodriguez}, \bibinfo{person}{Armand Joulin}, \bibinfo{person}{Edouard Grave}, {and} \bibinfo{person}{Guillaume Lample}.} \bibinfo{year}{2023}\natexlab{}.
\newblock \bibinfo{title}{LLaMA: Open and Efficient Foundation Language Models}.
\newblock
\newblock
\showeprint[arxiv]{2302.13971}~[cs.CL]
\urldef\tempurl%
\url{https://arxiv.org/abs/2302.13971}
\showURL{%
\tempurl}


\bibitem[von Ahn and Dabbish(2008)]%
        {vonahn2008designing}
\bibfield{author}{\bibinfo{person}{Luis von Ahn} {and} \bibinfo{person}{Laura Dabbish}.} \bibinfo{year}{2008}\natexlab{}.
\newblock \showarticletitle{Designing games with a purpose}. In \bibinfo{booktitle}{\emph{Communications of the ACM}}, Vol.~\bibinfo{volume}{51}. \bibinfo{pages}{58--67}.
\newblock


\bibitem[Wang et~al\mbox{.}(2025)]%
        {multichat}
\bibfield{author}{\bibinfo{person}{Bingbing Wang}, \bibinfo{person}{Yiming Du}, \bibinfo{person}{Bin Liang}, \bibinfo{person}{Zhixin Bai}, \bibinfo{person}{Min Yang}, \bibinfo{person}{Baojun Wang}, \bibinfo{person}{Kam-Fai Wong}, {and} \bibinfo{person}{Ruifeng Xu}.} \bibinfo{year}{2025}\natexlab{}.
\newblock \showarticletitle{A New Formula for Sticker Retrieval: Reply with Stickers in Multi-Modal and Multi-Session Conversation}. In \bibinfo{booktitle}{\emph{Proceedings of the AAAI Conference on Artificial Intelligence}}, Vol.~\bibinfo{volume}{39}. \bibinfo{pages}{25327--25335}.
\newblock


\bibitem[Wang et~al\mbox{.}(2024)]%
        {stickertag}
\bibfield{author}{\bibinfo{person}{Bingbing Wang}, \bibinfo{person}{Bin Liang}, \bibinfo{person}{Chun-Mei Feng}, \bibinfo{person}{Wangmeng Zuo}, \bibinfo{person}{Zhixin Bai}, \bibinfo{person}{Shijue Huang}, \bibinfo{person}{Kam-Fai Wong}, \bibinfo{person}{Xi Zeng}, {and} \bibinfo{person}{Ruifeng Xu}.} \bibinfo{year}{2024}\natexlab{}.
\newblock \showarticletitle{Towards Real-World Stickers Use: A New Dataset for Multi-Tag Sticker Recognition}.
\newblock \bibinfo{journal}{\emph{arXiv preprint arXiv:2403.05428}} (\bibinfo{year}{2024}).
\newblock


\bibitem[Wang et~al\mbox{.}(2022a)]%
        {wang2022git}
\bibfield{author}{\bibinfo{person}{Jiahui Wang} {et~al\mbox{.}}} \bibinfo{year}{2022}\natexlab{a}.
\newblock \showarticletitle{GIT: A Generative Image-to-text Transformer for Vision and Language}. In \bibinfo{booktitle}{\emph{CVPR}}.
\newblock


\bibitem[Wang et~al\mbox{.}(2023)]%
        {wang2023zero}
\bibfield{author}{\bibinfo{person}{Junjie Wang}, \bibinfo{person}{Ping Yang}, \bibinfo{person}{Ruyi Gan}, \bibinfo{person}{Yuxiang Zhang}, \bibinfo{person}{Jiaxing Zhang}, {and} \bibinfo{person}{Tetsuya Sakai}.} \bibinfo{year}{2023}\natexlab{}.
\newblock \showarticletitle{Zero-Shot Learners for Natural Language Understanding via a Unified Multiple-Choice Perspective}.
\newblock \bibinfo{journal}{\emph{IEEE Access}} (\bibinfo{year}{2023}).
\newblock


\bibitem[Wang et~al\mbox{.}(2022b)]%
        {wang2022ofa}
\bibfield{author}{\bibinfo{person}{Peng Wang} {et~al\mbox{.}}} \bibinfo{year}{2022}\natexlab{b}.
\newblock \showarticletitle{Unifying Architectures, Tasks, and Modalities Through a Simple Sequence-to-Sequence Learning Framework}. In \bibinfo{booktitle}{\emph{ICLR}}.
\newblock


\bibitem[Wilkinson et~al\mbox{.}(2016)]%
        {fair_principles}
\bibfield{author}{\bibinfo{person}{Mark~D. Wilkinson}, \bibinfo{person}{Michel Dumontier}, \bibinfo{person}{Ilja~J. Aalbersberg}, \bibinfo{person}{Gwen Appleton}, \bibinfo{person}{M. Axton}, \bibinfo{person}{Arie Baak}, {and} \bibinfo{person}{et al.}} \bibinfo{year}{2016}\natexlab{}.
\newblock \showarticletitle{The FAIR Guiding Principles for scientific data management and stewardship}.
\newblock \bibinfo{journal}{\emph{Scientific Data}}  \bibinfo{volume}{3} (\bibinfo{year}{2016}), \bibinfo{pages}{160018}.
\newblock


\bibitem[Yuan et~al\mbox{.}(2025)]%
        {vsd2m}
\bibfield{author}{\bibinfo{person}{Zhiqiang Yuan}, \bibinfo{person}{Jiapei Zhang}, \bibinfo{person}{Ying Deng}, \bibinfo{person}{Yeshuang Zhu}, \bibinfo{person}{Jie Zhou}, {and} \bibinfo{person}{Jinchao Zhang}.} \bibinfo{year}{2025}\natexlab{}.
\newblock \bibinfo{title}{VSD2M: A Large-scale Vision-language Sticker Dataset for Multi-frame Animated Sticker Generation}.
\newblock
\newblock
\showeprint[arxiv]{2412.08259}~[cs.HC]
\urldef\tempurl%
\url{https://arxiv.org/abs/2412.08259}
\showURL{%
\tempurl}


\bibitem[Zhang et~al\mbox{.}(2021)]%
        {zhang2021vinvl}
\bibfield{author}{\bibinfo{person}{Pengchuan Zhang} {et~al\mbox{.}}} \bibinfo{year}{2021}\natexlab{}.
\newblock \showarticletitle{VinVL: Revisiting Visual Representations in Vision-Language Models}. In \bibinfo{booktitle}{\emph{CVPR}}.
\newblock


\bibitem[Zhang et~al\mbox{.}(2020)]%
        {zhang2019bertscore}
\bibfield{author}{\bibinfo{person}{Tianyi Zhang}, \bibinfo{person}{Varsha Kishore}, \bibinfo{person}{Felix Wu}, \bibinfo{person}{Kilian~Q Weinberger}, {and} \bibinfo{person}{Yoav Artzi}.} \bibinfo{year}{2020}\natexlab{}.
\newblock \showarticletitle{BERTScore: Evaluating Text Generation with BERT}. In \bibinfo{booktitle}{\emph{International Conference on Learning Representations (ICLR)}}.
\newblock
\urldef\tempurl%
\url{https://openreview.net/forum?id=SkeHuCVFDr}
\showURL{%
\tempurl}


\bibitem[Zhang et~al\mbox{.}(2024)]%
        {zhang2024stickerconv}
\bibfield{author}{\bibinfo{person}{Yiqun Zhang}, \bibinfo{person}{Fanheng Kong}, \bibinfo{person}{Peidong Wang}, \bibinfo{person}{Shuang Sun}, \bibinfo{person}{Lingshuai Wang}, \bibinfo{person}{Shi Feng}, \bibinfo{person}{Daling Wang}, \bibinfo{person}{Yifei Zhang}, {and} \bibinfo{person}{Kaisong Song}.} \bibinfo{year}{2024}\natexlab{}.
\newblock \showarticletitle{StickerConv: Generating Multimodal Empathetic Responses from Scratch}.
\newblock \bibinfo{journal}{\emph{arXiv preprint arXiv:2402.01679}} (\bibinfo{year}{2024}).
\newblock


\bibitem[Zhao et~al\mbox{.}(2023)]%
        {stickerclip}
\bibfield{author}{\bibinfo{person}{Sijie Zhao}, \bibinfo{person}{Yixiao Ge}, \bibinfo{person}{Zhongang Qi}, \bibinfo{person}{Lin Song}, \bibinfo{person}{Xiaohan Ding}, \bibinfo{person}{Zehua Xie}, {and} \bibinfo{person}{Ying Shan}.} \bibinfo{year}{2023}\natexlab{}.
\newblock \showarticletitle{Sticker820k: Empowering interactive retrieval with stickers}.
\newblock \bibinfo{journal}{\emph{arXiv preprint arXiv:2306.06870}} (\bibinfo{year}{2023}).
\newblock


\bibitem[Zhaoolee(2024)]%
        {chinesebqb}
\bibfield{author}{\bibinfo{person}{Zhaoolee}.} \bibinfo{year}{2024}\natexlab{}.
\newblock \bibinfo{title}{ChineseBQB}.
\newblock
\newblock
\urldef\tempurl%
\url{https://github.com/zhaoolee/ChineseBQB}
\showURL{%
\tempurl}
\newblock
\shownote{Accessed: 2025-05-27}.


\bibitem[Zhu et~al\mbox{.}(2023)]%
        {zhu2023minigpt}
\bibfield{author}{\bibinfo{person}{Deyao Zhu}, \bibinfo{person}{Jun Chen}, \bibinfo{person}{Xiaoqian Shen}, \bibinfo{person}{Xiang Li}, {and} \bibinfo{person}{Mohamed Elhoseiny}.} \bibinfo{year}{2023}\natexlab{}.
\newblock \showarticletitle{Minigpt-4: Enhancing vision-language understanding with advanced large language models}.
\newblock \bibinfo{journal}{\emph{arXiv preprint arXiv:2304.10592}} (\bibinfo{year}{2023}).
\newblock


\end{thebibliography}

\appendix
\onecolumn










\end{document}